\documentclass{JINST}

\title{Performance of the RPC-based ALICE muon trigger system at the LHC}

\author{F. Boss\`u$^a$\thanks{Now at iThemba LABS, Somerset West, South Africa.}, M. Gagliardi$^a$\thanks{Corresponding
author.}~ and M. Marchisone$^{a,b}$, for the ALICE Collaboration\\
\llap{$^a$}Universit\`a degli Studi and Sezione INFN di Torino,\\
  Via Giuria 1, 10125 Torino, Italy\\
\llap{$^b$}Universit\'e B. Pascal and LPC Clermont-Ferrand, CNRS-IN2P3, \\  24 Avenue des Landais, 63171 Aubi\`ere Cedex, France\\
  E-mail: \email{Martino.Gagliardi@cern.ch}}

\abstract{The forward muon spectrometer of ALICE (A Large Ion Collider Experiment) is equipped with a trigger system made of four planes of Resistive Plate Chambers (RPC), arranged in two stations with two planes each, for a total area of about 140~m$^2$. The system  provides single and di-muon triggers with suitable transverse momentum selection, optimised for the physics of quarkonia and open heavy flavour.
In the first two years of data-taking at the Large Hadron Collider (2010 and 2011) the 72 RPCs were operated in highly saturated avalanche mode in both pp and Pb--Pb collisions. The integrated charge was about 1.3~mC/cm$^{2}$ on average  and 3.5~mC/cm$^{2}$ for the most exposed detectors.
This paper describes two main results. The first result is the determination of the RPC performance, with particular focus on the stability of the main detector parameters such as efficiency, dark current, and dark rate. The second result is the measurement of the muon trigger performance in Pb--Pb collisions at $\sqrt{s_{NN}}$~=~2.76~TeV, in terms of the reliability and stability of the trigger decision logic.
\vspace{1 cm}

We dedicate this piece of work to the memory of Anna Piccotti.
}

\vspace{1 cm}


\begin{document}

\section{Introduction}

%\subsection{The ALICE muon trigger system}

ALICE (A Large Ion Collider Experiment~\cite{aliJINST}) studies nuclear matter at very high temperatures and energy densities, and the transition to a deconfined partonic phase, known as Quark Gluon Plasma~\cite{karsch} (QGP). This is done via the analysis of ultra-relativistic heavy-ion collisions at the Large Hadron Collider (LHC). Proton-proton physics is also included in the ALICE program, both as a reference for observables measured in heavy-ion collisions and as a \textit{per se} field of study. 

Heavy flavour production is sensitive to the properties of QGP. In the ALICE forward muon spectrometer, heavy flavoured mesons are detected via their muonic and semi-muonic decays, in both Pb--Pb~\cite{hfMuon, RAA,suire} and pp~\cite{jpsi7TeV, jpsi276TeV, jpsiPol,singleMuon7TeV} collisions.  The spectrometer is composed  by a set of absorbers,  a muon tracking system, a dipole magnet and a muon trigger system, whose task is to identify muons and to reduce the background of low transverse momentum ($p_{\rm{T}}$) muons from light hadron decays. The muon trigger system consists of 72 Resistive Plate Chamber~\cite{santCard} (RPC) modules arranged in two stations, located, respectively, at a distance of 16~m and 17~m from the interaction point. Each station is made of two detection planes with 18 RPCs each. The detection planes are arranged perpendicular to the beam line.  The total active area of the system is about 140~m$^2$. The total number of electronics channels is about 21000. The spatial information provided  (with subcentimeter resolution~\cite{spatRes}) by the RPCs is used to perform a selection on the muon $p_{\rm{T}}$,  via the deviation with respect to the trajectory of an infinite momentum track originated at the interaction point. The system is able to deliver single and di-muon (unlike- and like-sign) triggers. For each of these signals, two different $p_{\rm{T}}$ thresholds can be handled simultaneously, for a total of six trigger signals evaluated and delivered to the ALICE trigger processor at a frequency of 40~MHz and with a latency of about 800~ns. The first-level muon trigger decision is performed by a set of 234 electronics boards. More details about the trigger algorithm and electronics are given in~\cite{yermia}. The fraction of operational Front-End Electronics (FEE) channels was 99.7\% at the end of the 2011 data-taking.  

%\subsection{Operating conditions in 2010-11} \label{subsec:opCond}

In 2010 and 2011, ALICE took pp collision data at $\sqrt{s}$~=~7~TeV (8~months/year)\footnote{In 2011, data were also taken for a few days in pp collisions at $\sqrt{s}$~=~2.76~TeV.} and in Pb--Pb collisions at $\sqrt{s_{NN}}$~=~2.76 TeV (1~month/year). 
In 2010, the typical luminosity in pp collisions was about 10$^{29}$~cm$^{-2}$s$^{-1}$, corresponding to a single muon trigger rate of about 100~Hz, with a $p_{\rm{T}}$ threshold of 0.5~GeV/$c$. The luminosity in Pb--Pb collisions was about 10$^{25}$~cm$^{-2}$s$^{-1}$. Given the low luminosity, no dedicated muon trigger was used in Pb--Pb collisions; however, the muon trigger data were read out and used for offline analysis. 
In 2011, the luminosity in pp collisions was about 2$\times$10$^{30}$~cm$^{-2}$s$^{-1}$, corresponding to a single muon trigger rate of about 500~Hz and to a di-muon trigger rate of about 20~Hz, with a $p_{\rm{T}}$ threshold of 1~GeV/$c$.  The luminosity in Pb--Pb collisions was about 3$\times$10$^{26}$~cm$^{-2}$s$^{-1}$, corresponding to a single muon trigger rate of about 500~Hz and to a di-muon trigger rate of about 200~Hz, with a $p_{\rm{T}}$ threshold of 1~GeV/$c$.

\section{Performance and stability of the ALICE muon trigger RPCs}

%\subsection{Main features and operating parameters}

The ALICE muon trigger detectors are 2~mm single gap RPCs, with low resistivity ($\simeq$10$^9$~$\Omega$cm) bakelite electrodes. They are operated in highly saturated avalanche mode~\cite{rpc2005}, with a gas mixture consisting of 89.7\%~C$_2$H$_2$F$_4$, 10\%~C$_4$H$_{10}$, 0.3\%~SF$_6$. The gas relative humidity is kept at 37\%, in order to prevent alterations in the bakelite resistivity~\cite{rpc2010, hum}.  The individual RPC areas range from (72$\times$223)~cm$^2$ to (76$\times$292)~cm$^2$.
The signal is picked up inductively on both sides of the detector by means of orthogonal copper strips, in the X (bending plane) and Y (non-bending plane) directions. Strips with pitch of 1~cm, 2~cm and 4~cm and length ranging from 17~cm to 72~cm are employed; areas closer to the beam line have the finest segmentation. The RPC signal is discriminated in the FEE without pre-amplification \cite{adult}. The signal amplitude threshold of the FEE can be separately adjusted for each RPC via the Detector Control System (DCS); it is set to 7~mV for most RPCs.  

The operating high voltage was optimised for each RPC with cosmic data, and fine-tuned with early pp collision data~\cite{rpc2010,rpc2007}: the chosen values range from 10~kV to 10.4~kV. Operating voltage correction to compensate for temperature and pressure variations is performed online by the DCS.

In 2010 and 2011 the detectors were exposed to about 13~Mhit/cm$^2$ on average and 35~Mhit/cm$^2$ for the most exposed RPCs, corresponding to an integrated charge of 1.3~mC/cm$^2$ on average and 3.5~mC/cm$^2$ for the most exposed RPCs (as will be shown in~\ref{countRateCurr}, the average charge per hit is about 100~pC). Most of the hits were integrated in 2011.  RPC prototypes operated with the same gas mixture have been ageing-tested~\cite{rpc2005, tesiPoggio} up to an exposure of 550~Mhit/cm$^2$, corresponding to about ten years of safe operation at the LHC in the expected running conditions. 

%\subsection{RPC performance and stability}    

\subsection{Efficiency}

The RPC efficiency can be measured from data: since the trigger algorithm requires tracks to have hits in three out of four detection planes, the efficiency of a detection element in a given plane can be obtained by using the remaining three planes as an external tracking system~\cite{stocco}.

\begin{figure}[!hbtp]
   \centering
          \includegraphics[width=0.4\textwidth]{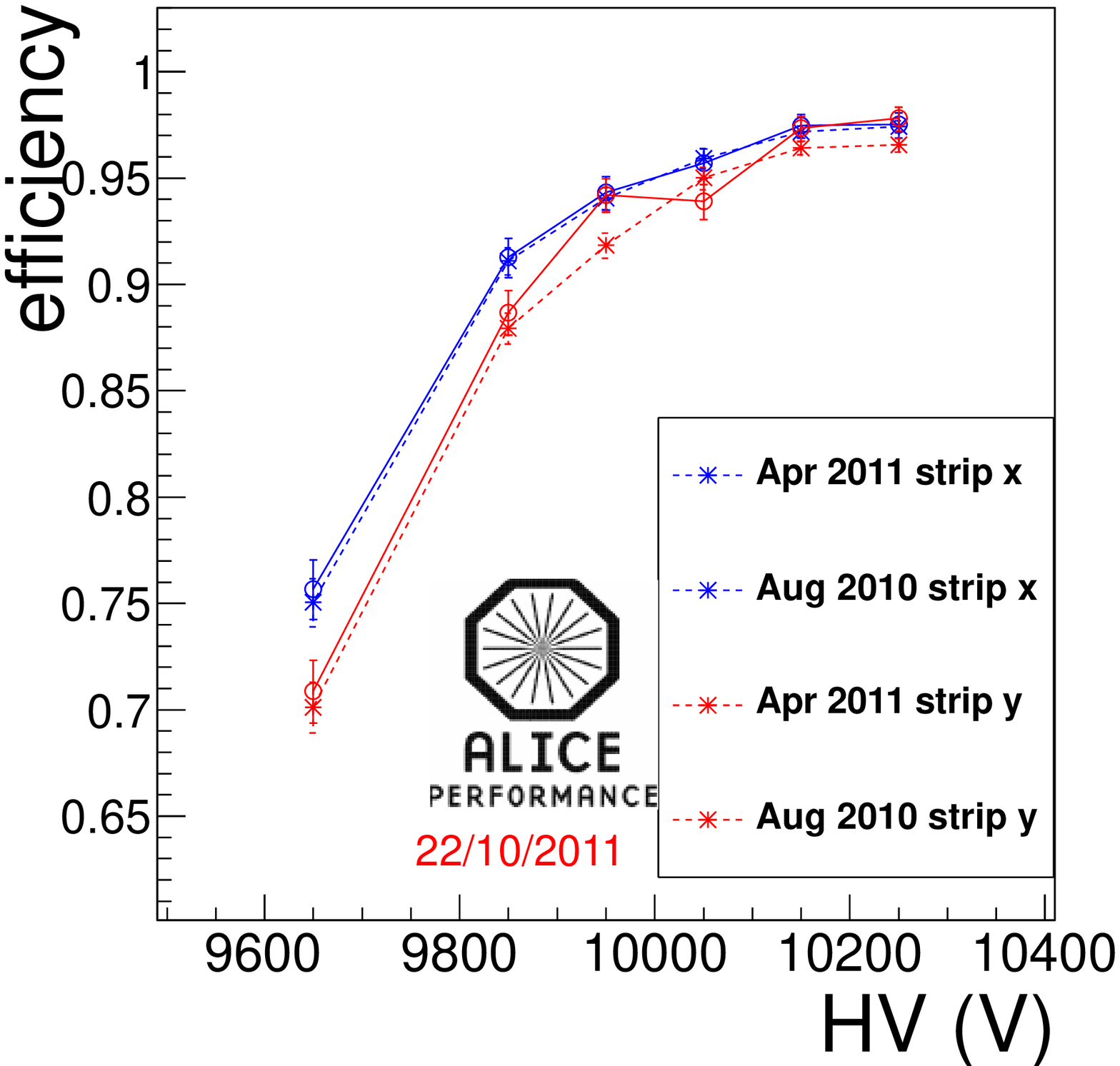}
	  \includegraphics[width=0.55\textwidth]{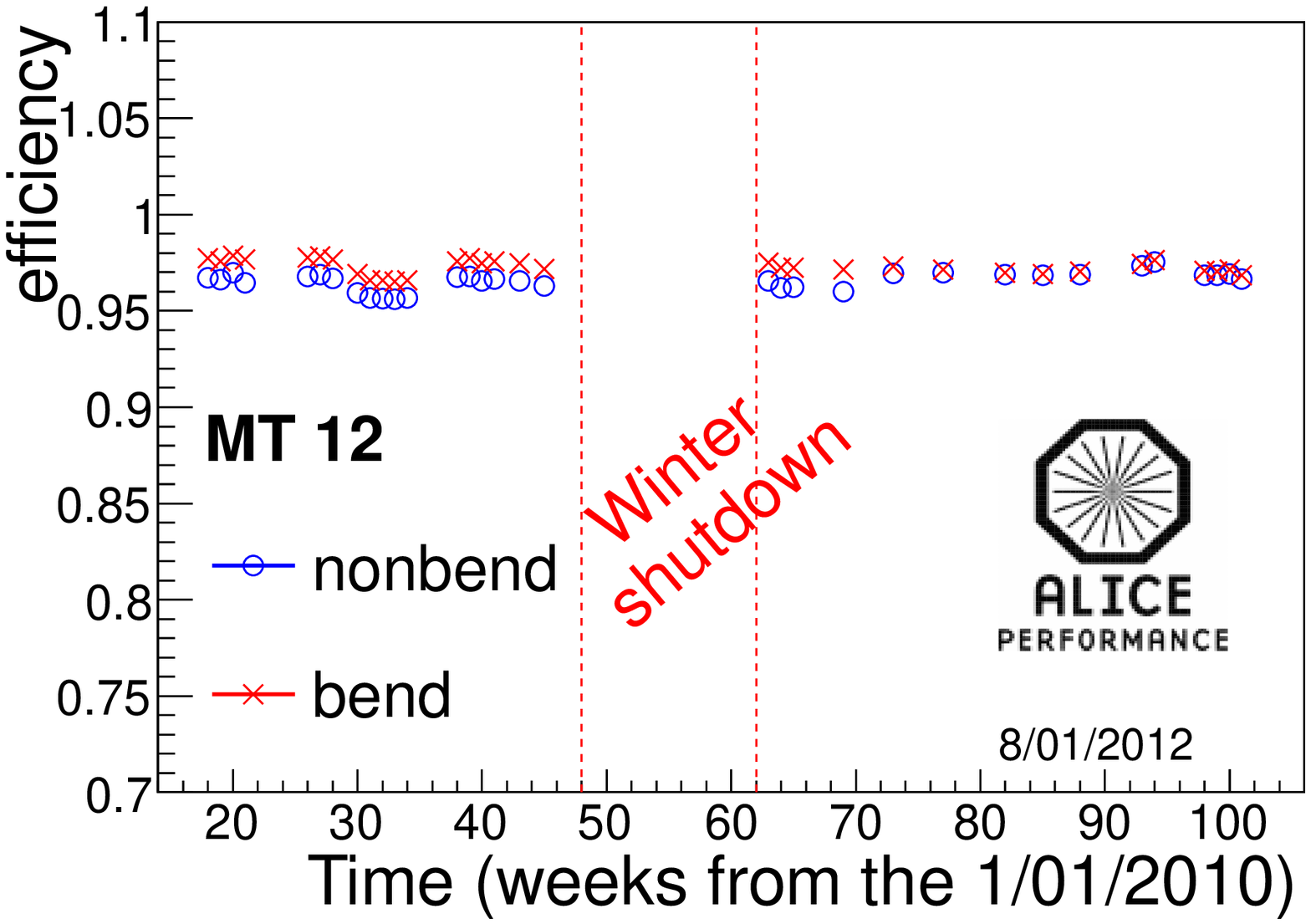}
            \caption{ Left: efficiency curve of one RPC, as measured with pp collisions in two different high voltage scans 8 months apart. High voltage values are corrected by temperature and pressure variations. Right: average efficiency of one of the four detection planes (18 RPCs), as a function of time, in 2010 and 2011. The efficiency is measured separately for the X (bending) and Y (non-bending) planes. }
   \label{fig:eff}
\end{figure}

In order to check the stability of the efficiency plateau, two high voltage scans were performed within 8 months from one another. A very good reproducibility of the efficiency curve was observed  (an example is shown in figure~\ref{fig:eff}, left). The measured shift of the working point was less than 50~V ($\simeq$0.5\%) for all RPCs.

The RPC efficiency is constantly monitored in order to provide efficiency maps for offline analysis and to check its stability in time. The results are summarised in figure~\ref{fig:eff}, right: the average RPC efficiency, about 95\%, has been stable within 0.5\% in two years of operation. 

\subsection{Cluster size}

\begin{figure}[!hbtp]
   \centering
           \includegraphics[width=0.55\textwidth, height=0.35\textwidth]{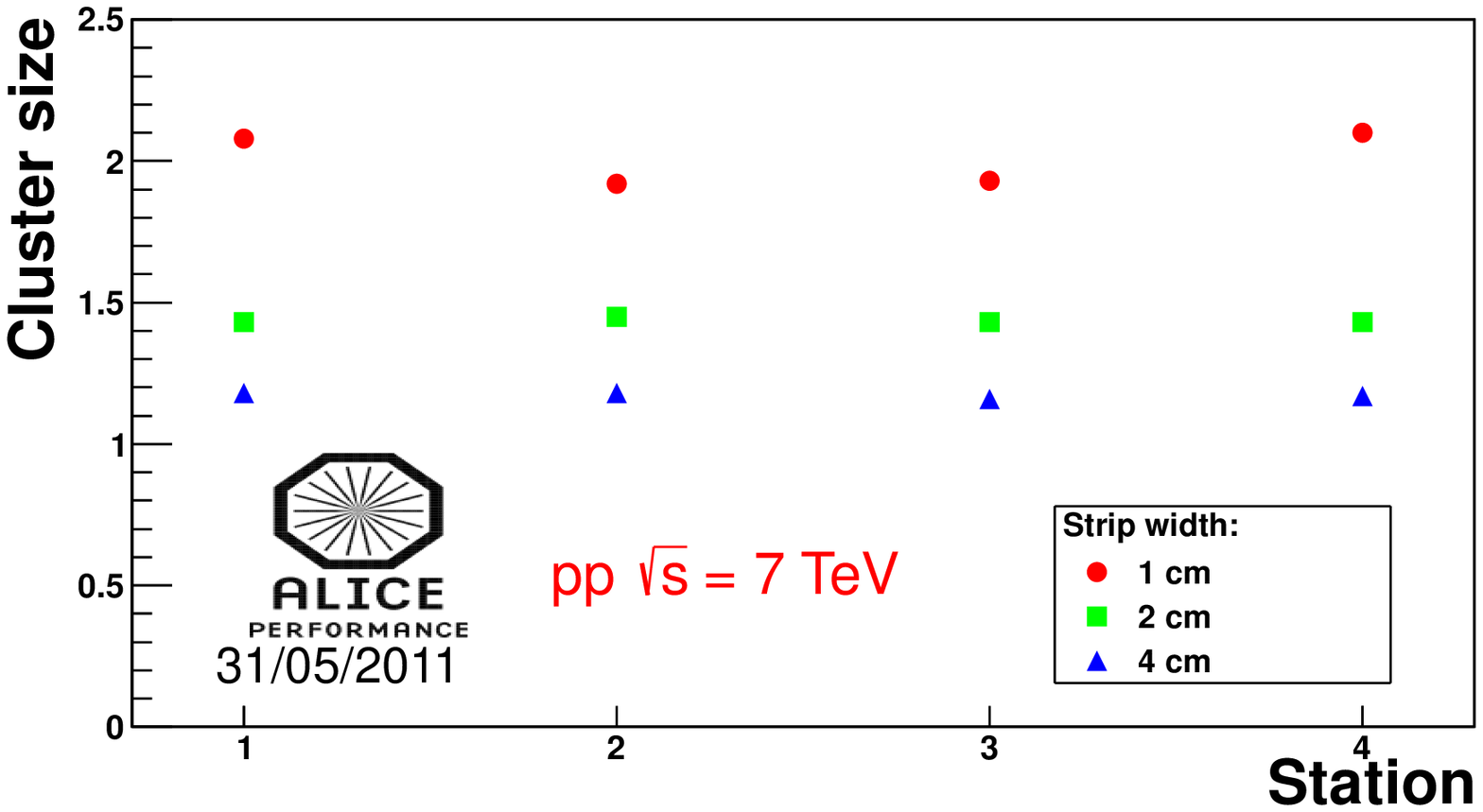}
	   \includegraphics[width=0.55\textwidth, height=0.35\textwidth]{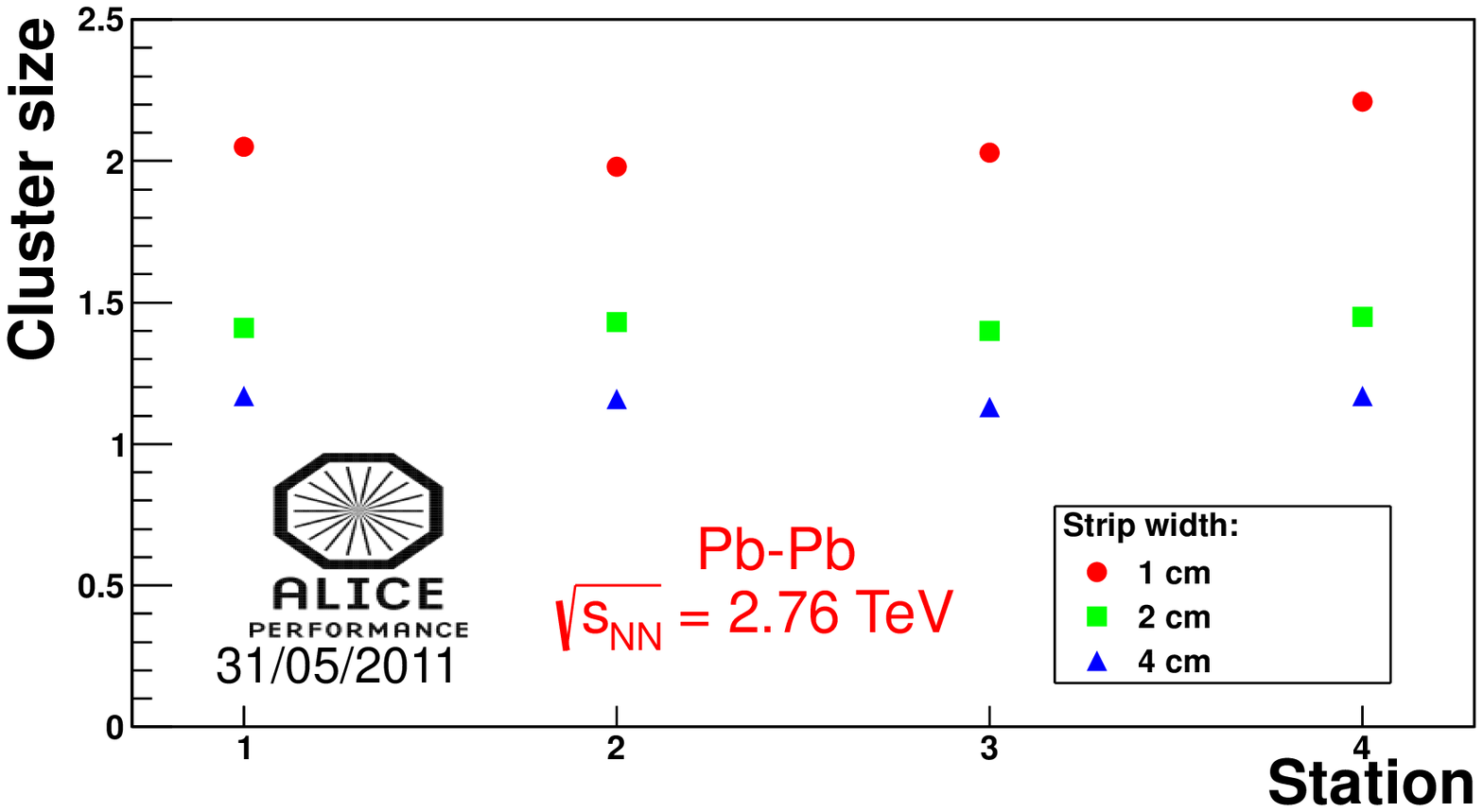}
       \caption{Average RPC cluster size for the four detection planes (18 RPC/plane), for strips with pitch 1~cm, 2~cm and 4~cm, in pp collisions at $\sqrt{s}$~=~7~TeV (top) and in Pb--Pb collisions at $\sqrt{s_{NN}}$~=~2.76~TeV (bottom).}
   \label{fig:cluSize}    
\end{figure}

The RPC cluster size was measured in both pp and Pb--Pb collisions. The results obtained with 2010 data are shown in figure~\ref{fig:cluSize} for the three different strip pitches employed in the RPCs.

The measured average value for strips of 2~cm is 1.40, in good agreement with the value of 1.33 measured in beam tests~\cite{rpc2005} during the R\&D phase. The slight difference is compatible with the fact that the FEE threshold was set to 10~mV in the beam test and 7~mV in the current setup. No significant difference was found between the cluster size measured in pp and Pb--Pb collisions. The same analysis was performed with 2011 data, with compatible results.

%\subsection{Dark rate and current}

\subsection{Counting rate and current} \label{countRateCurr}

\begin{figure}[!hbtp]
   \centering
           \includegraphics[width=0.55\textwidth]{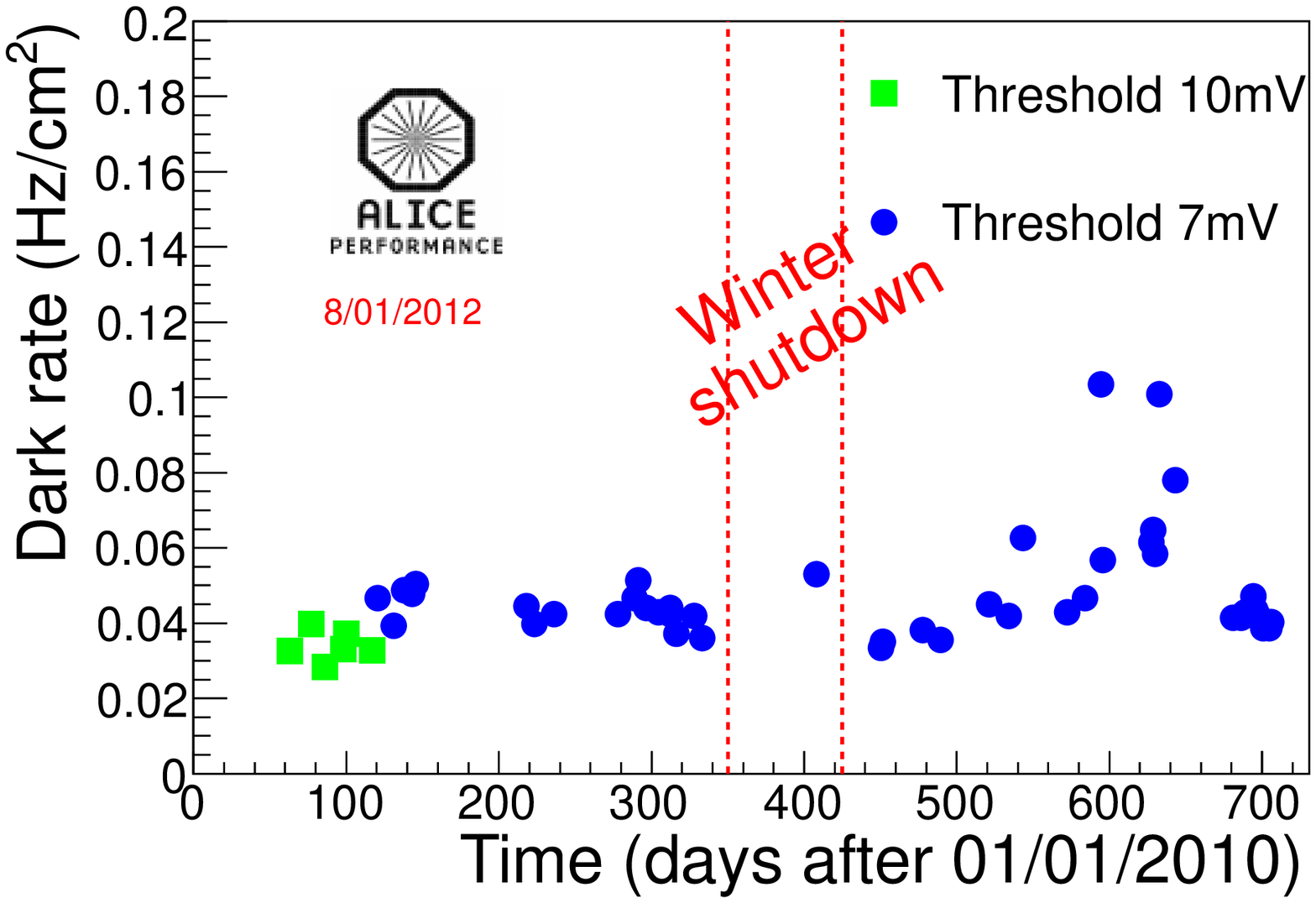}
	   \includegraphics[width=0.55\textwidth]{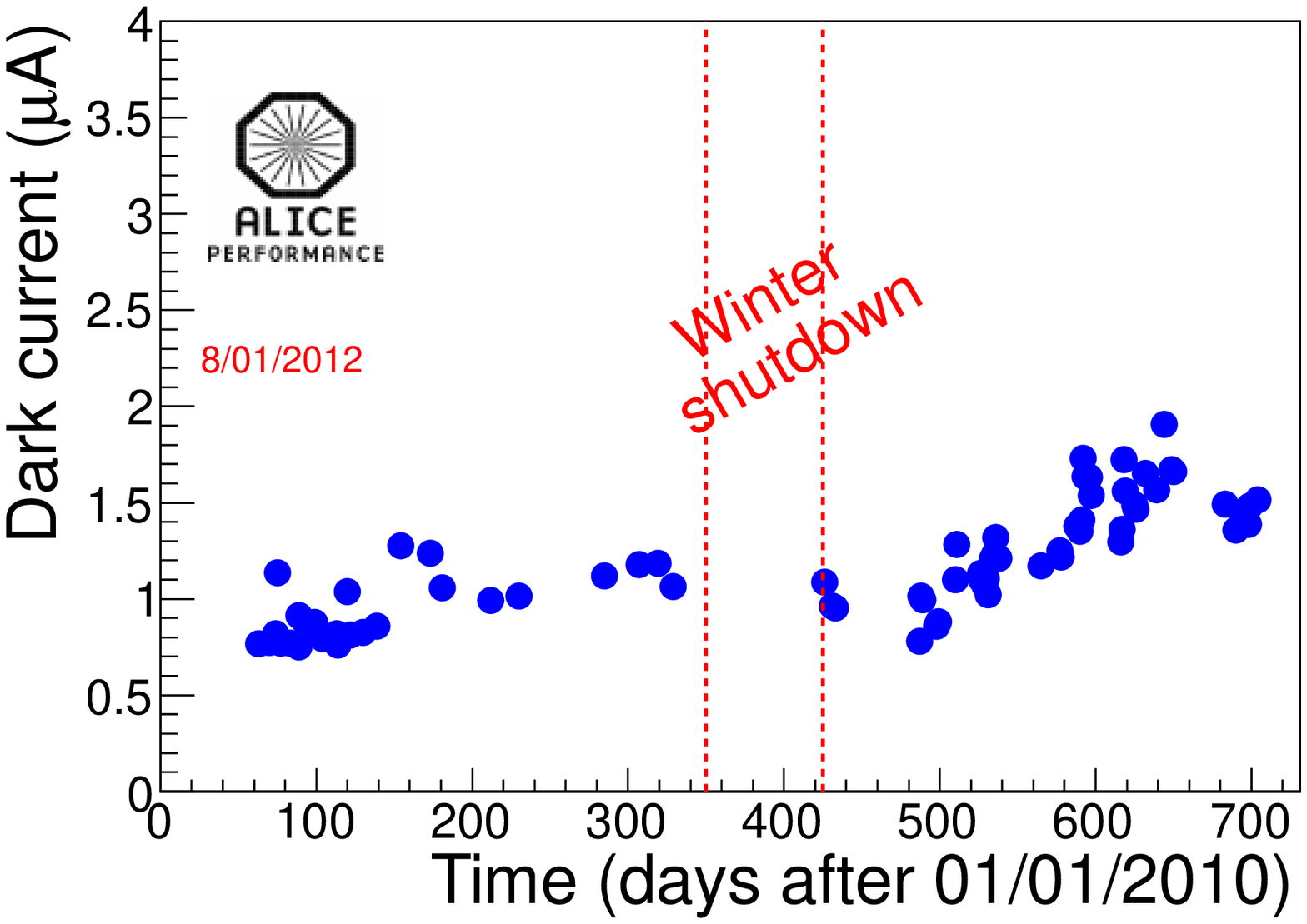}
          \caption{Average RPC dark rate (top) and dark current (bottom) as a function of time in 2010 and 2011.}
   \label{fig:dark} 
\end{figure}

The RPC dark current and counting rate were periodically monitored. The results are depicted in figure~\ref{fig:dark}. The dark counting rate was measured from scalers, in dedicated runs taken right after the physics fills. Its average value of 0.05~Hz/cm$^2$ is very stable in time. The spikes seen in the autumn of 2011 can be ascribed to beam-induced afterglow, since in this period the LHC proton beams reached the highest intensity \cite{mfl}; in the following Pb--Pb run (November 2011), with much lower intensities, the dark rate is again very stable. The average dark current shows a slightly increasing trend, which seems to be temporarily inverted or mitigated by long periods with high voltage turned off (e.g. the winter shutdown); the average value is 1.5~$\mu$A, corresponding to about 0.1~nA/cm$^2$. 

The maximum average current and counting rate reached during data-taking were measured to be about 1~nA/cm$^2$ and 10~Hz/cm$^2$, respectively. In such conditions (autumn 2011), the RPC currents and rates are dominated by the machine-induced background \cite{bossu}. Beam tests \cite{rpc2005, tesiPoggio} have shown that the RPC performance are unaffected up to rates of about 80~Hz/cm$^2$.

\begin{figure}[!hbtp]
   \centering
           \includegraphics[width=0.55\textwidth]{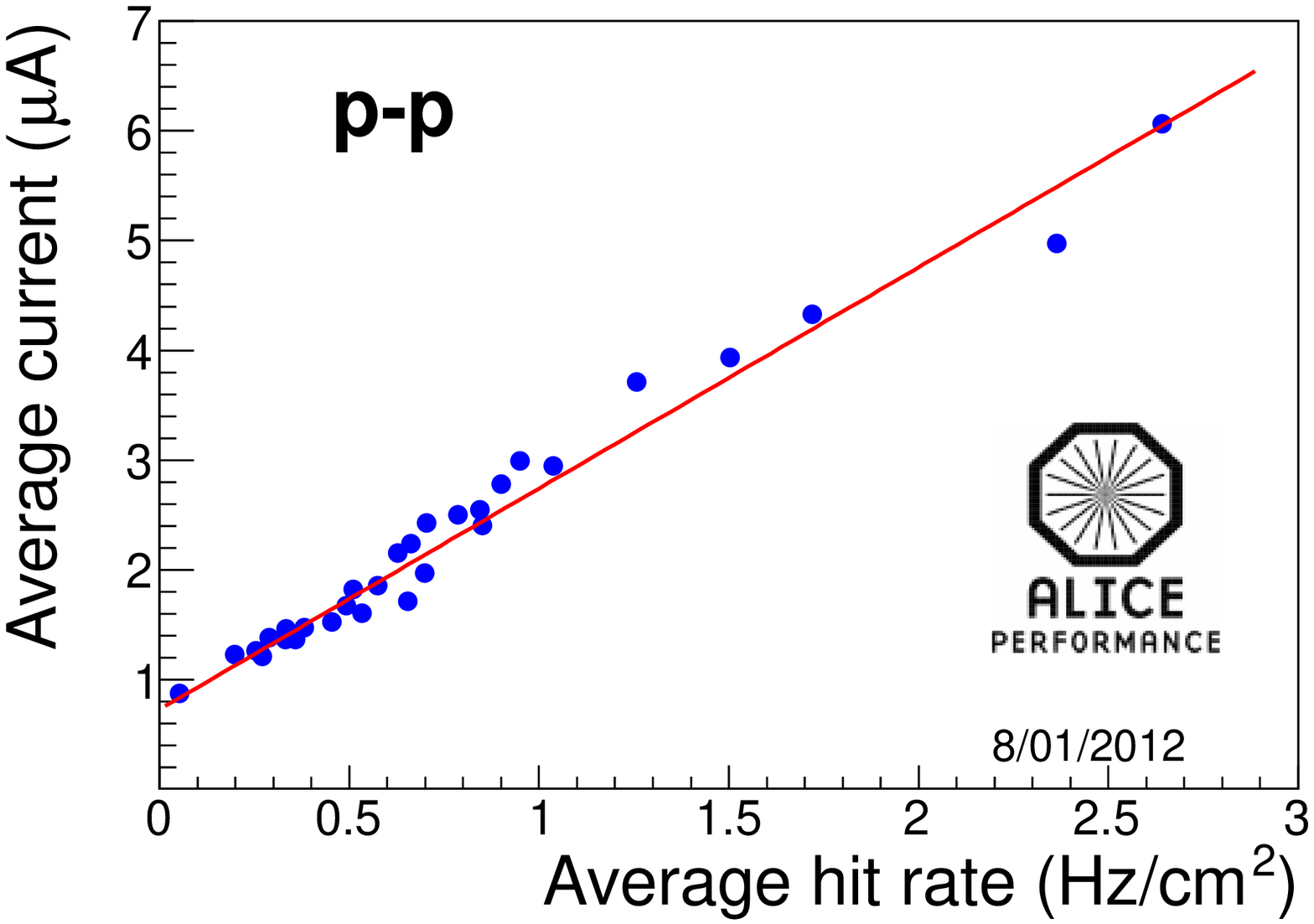}
	   \includegraphics[width=0.55\textwidth]{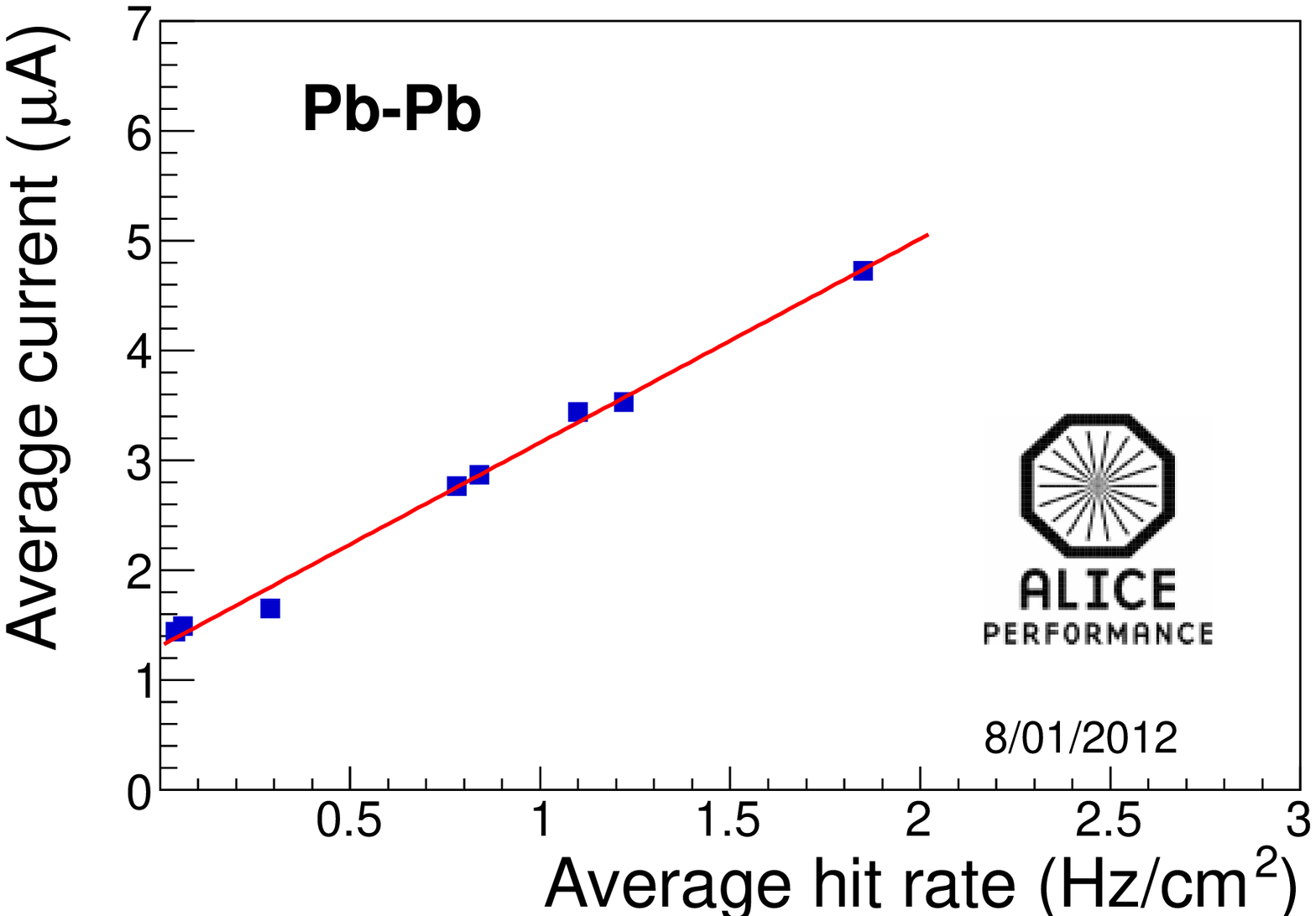}
         \caption{RPC average current as a function of the  average counting rate in pp (top) and Pb--Pb (bottom) collisions, with superimposed linear fit. The fitted slopes are 2~$\mu$A/(Hz/cm$^2$) for pp collisions and 1.9~$\mu$A/(Hz/cm$^2$) for Pb--Pb collisions.}
   \label{fig:IvsR}  
\end{figure}

In figure~\ref{fig:IvsR} the average RPC current during physics data-taking is plotted as a function of the average counting rate: as expected, a linear correlation is found. The slope of the curve is about 2~$\mu$A/(Hz/cm$^2$), corresponding to an average charge per hit of about 100~pC.

\begin{figure}[!hbtp]
   \centering
           \includegraphics[width=0.55\textwidth]{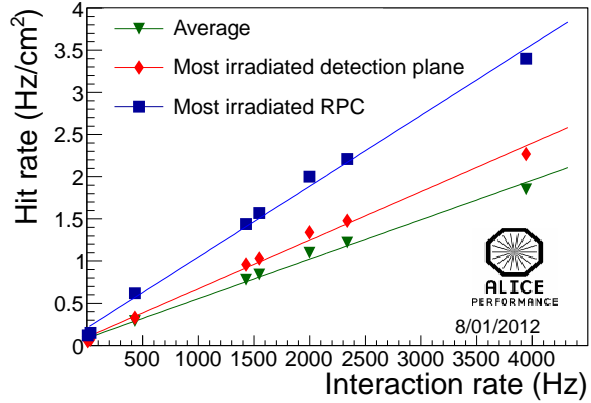}   
            \caption{RPC counting rate as a function of the interaction rate in Pb--Pb collisions: for the most exposed RPC (blue squares); average for the RPCs on the most exposed detection plane (red dots); average for all RPCs (green triangles). Lines are linear fits to the data.}
   \label{fig:RvsLumi}
\end{figure}

Figure~\ref{fig:RvsLumi} shows the RPC counting rate as a function of the minimum bias trigger rate in Pb--Pb collisions. The three curves correspond to the most exposed RPC, to the average of the RPCs on the most exposed detection plane and to the average of all RPCs. The slope of the curve corresponds to the average number of hits per minimum bias event: this quantity is 0.5$\times$10$^3$~cm$^{-2}$ on average and 0.8$\times$10$^3$~cm$^{-2}$ for the most exposed RPC. An extrapolation to 50~kHz minimum bias rate (hypothetical scenario for an upgraded LHC \cite{rateUpgrade}) leads to counting rates of 25~Hz/cm$^2$ on average and 40~Hz/cm$^2$ for the most exposed RPC. Such rates are still tolerable by the detectors in terms of rate capability. However, the detector lifetime in such a scenario might be limited: thus, the possibility of switching to a lower-gain gas mixture and a new FEE with amplification in order to reduce ageing effects is being considered.

\section{Performance of the ALICE muon trigger system in Pb--Pb collisions}

In this section the measured muon trigger multiplicities in Pb--Pb collisions and the performance of the trigger algorithm in 2010 and 2011 are described.

 %The results have been obtained with the sets of thresholds shown in table \ref{ptcut}.

%\begin{table}[htbp]
%\begin{center}
%\caption{Muon p$_{\rm{T}}$ thresholds used in Pb--Pb collisions}
%\label{ptcut}
%\begin{tabular}{|c|c|c|}
%\cline{2-3}
%\multicolumn{1}{c|}{} & \textbf{low-p$_{\rm{T}}$ threshold} & \textbf{high-p$_{\rm{T}}$ threshold} \\
%\hline
%\textbf{Pb--Pb 2010}  & 0.5~GeV/$c$                 & 1~GeV/$c$                    \\
%\hline
%\textbf{Pb--Pb 2011}  & 1~GeV/$c$                   & 4~GeV/$c$                    \\
%\hline
%\end{tabular}
%\end{center}
%\end{table}

\subsection{Multiplicities from the trigger algorithm}\label{subsec:mult}

\begin{figure}[htbp]
\centering
\includegraphics[width=0.55\textwidth]{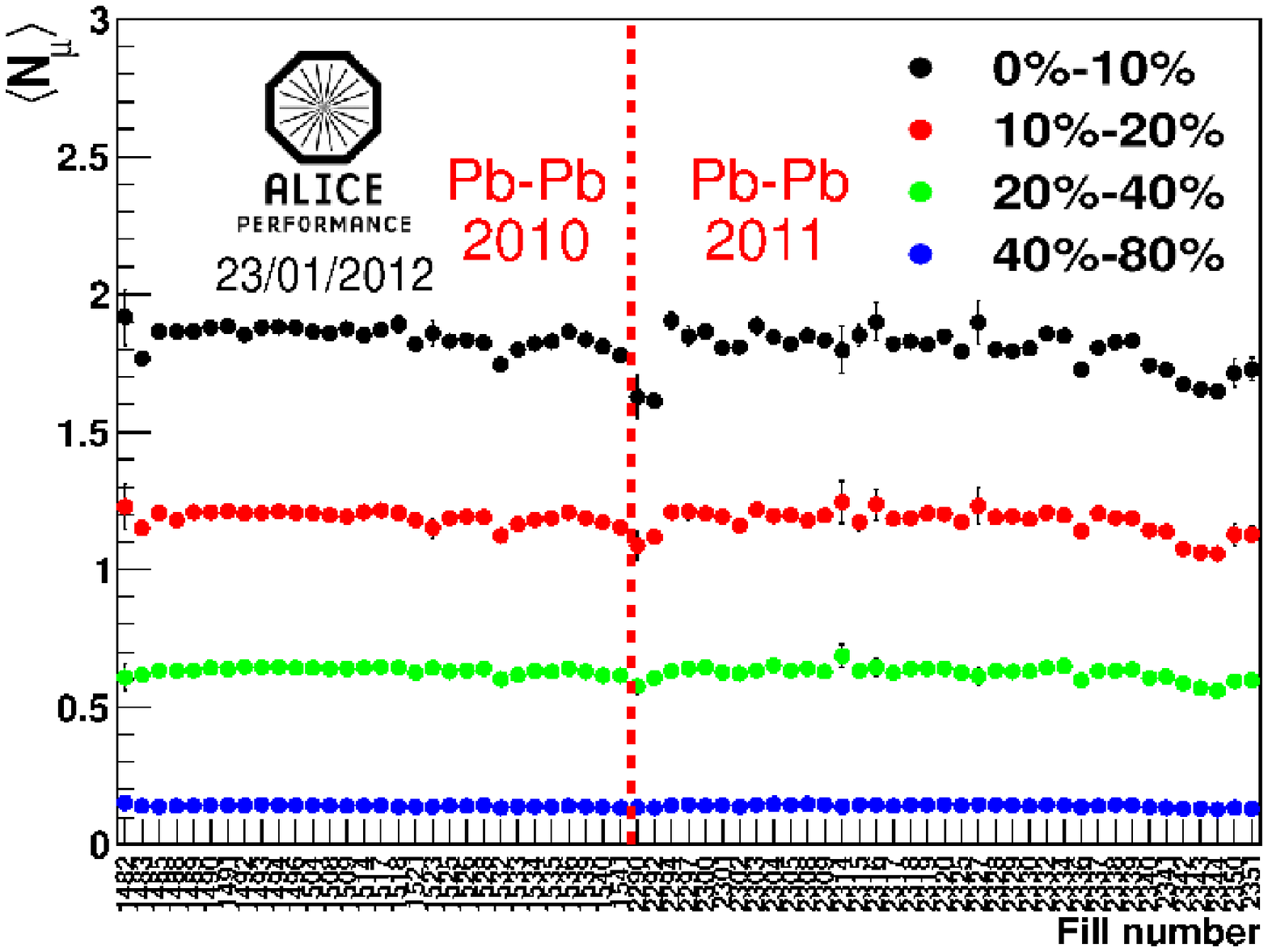}
\caption{Average number of muons from the trigger algorithm as a function of the fill number, for various Pb--Pb collision centrality bins.} 
\label{multTime}
\end{figure}

In figure \ref{multTime} the muon multiplicity (average number of muons detected by the trigger algorithm per Pb--Pb collision)  is shown for different centrality bins, ranging from 0-10\% (most central) to 40\%-80\% (most peripheral).  The centrality is measured as described in~\cite{prl106}. Muons are required to satisfy the trigger condition with the lowest possible $p_{\rm{T}}$ threshold ($p_{\rm{T}}\simeq$~0.5~GeV/$c$)  and to match a reconstructed track in the tracking system. The multiplicity is shown as a function of the LHC fill number, for a time span corresponding to about one month in both 2010 and 2011. 

As expected, we observe that the muon multiplicity increases with the Pb--Pb collision centrality, since so does the number of binary nucleon-nucleon collisions.  The stability of the detector response over time is satisfactory; the visible structures can be explained by slight changes in the number of active channels in the muon tracking system, affecting the muon reconstruction efficiency.

%With the same selection criteria, it is possible to calculate the average
%number of hit strips per minimum bias event in the same centrality bins.
%In figure \ref{multTime} (right), this is shown for the first trigger plane in
%the bending direction as a function of fill number.

%The average muon multiplicities for the four considered centrality classes are reported in table \ref{muonMult}.
%\begin{table}[htbp]
%\begin{center}
%\caption{Average number of muons as a function of centrality}
%\label{muonMult}
%\begin{tabular}{|c|c|c|c|}
%\hline
%\textbf{0\%--10\%} & \textbf{10\%--20\%} & \textbf{20\%--40\%} & \textbf{40\%--80\%} \\
%\hline
%1.83$\pm$0.16 & 1.18$\pm$0.10 & 0.62$\pm$0.07 & 0.14$\pm$0.01 \\
%\hline
%\end{tabular}
%\end{center}
%\end{table}

%\begin{table}[htbp]
%\begin{center}
%\caption{Average number of hit strips as a function of centrality (first trigger plane, bending direction)}
%\label{stripMult}
%\begin{tabular}{|c|c|c|c|}
%\hline
%\textbf{0\%--10\%} & \textbf{10\%--20\%} & \textbf{20\%--40\%} & \textbf{40\%--80\%} \\
%\hline
%2.93$\pm$0.39 & 1.86$\pm$0.24 & 0.97$\pm$0.12 & 0.21$\pm$0.02 \\
%\hline
%\end{tabular}
%\end{center}
%\end{table}

%In the analysis of the strip multiplicity soft background is not included: only
%strips participating in tracks
%recognized by the algorithm are taken into account. This means that the difference
%with the average number of muons is mostly due to the cluster size.

\subsection{Muon trigger turn-on curve}

\begin{figure}[htbp]
\centering
\includegraphics[width=0.55\textwidth]{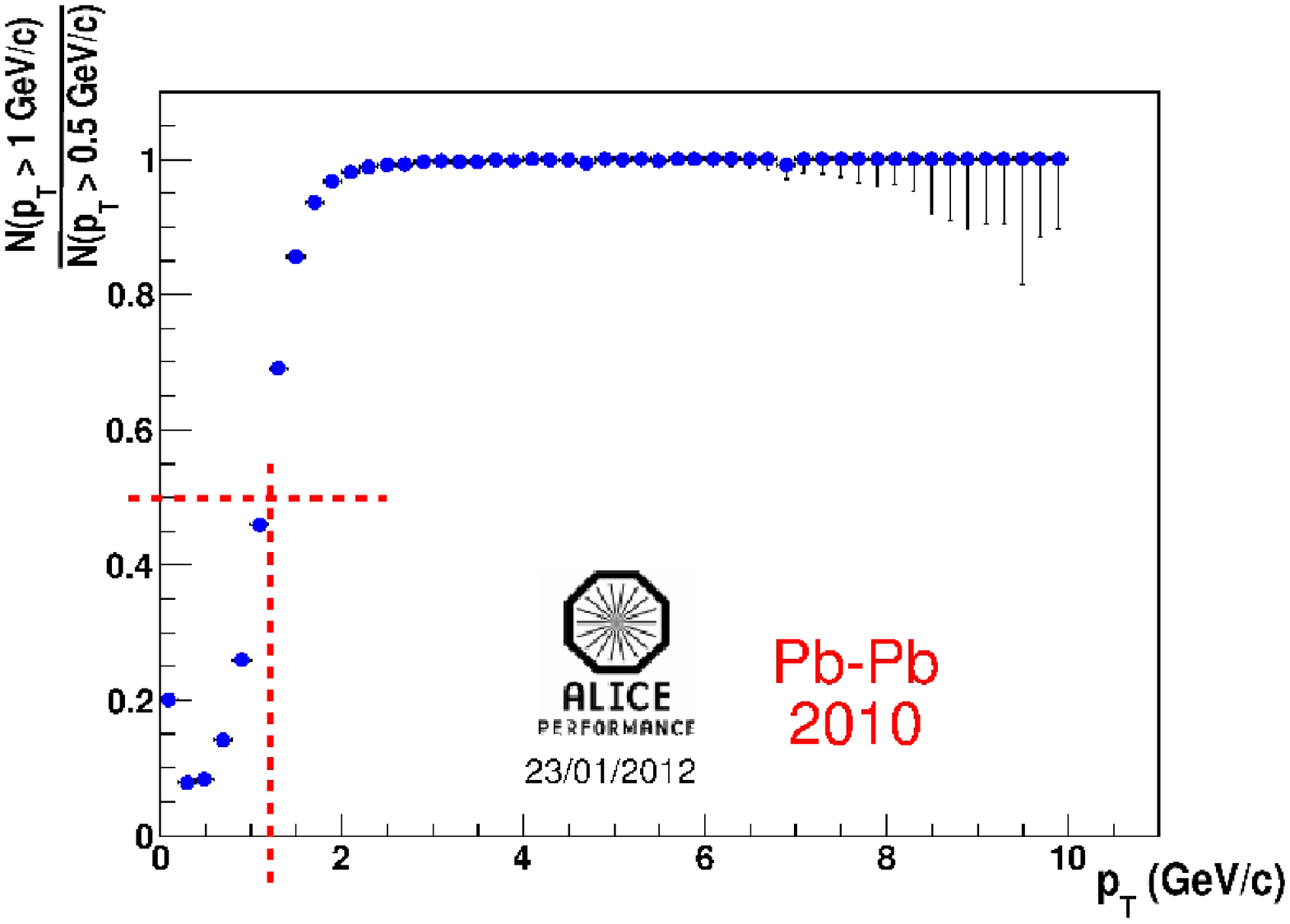}
\includegraphics[width=0.55\textwidth]{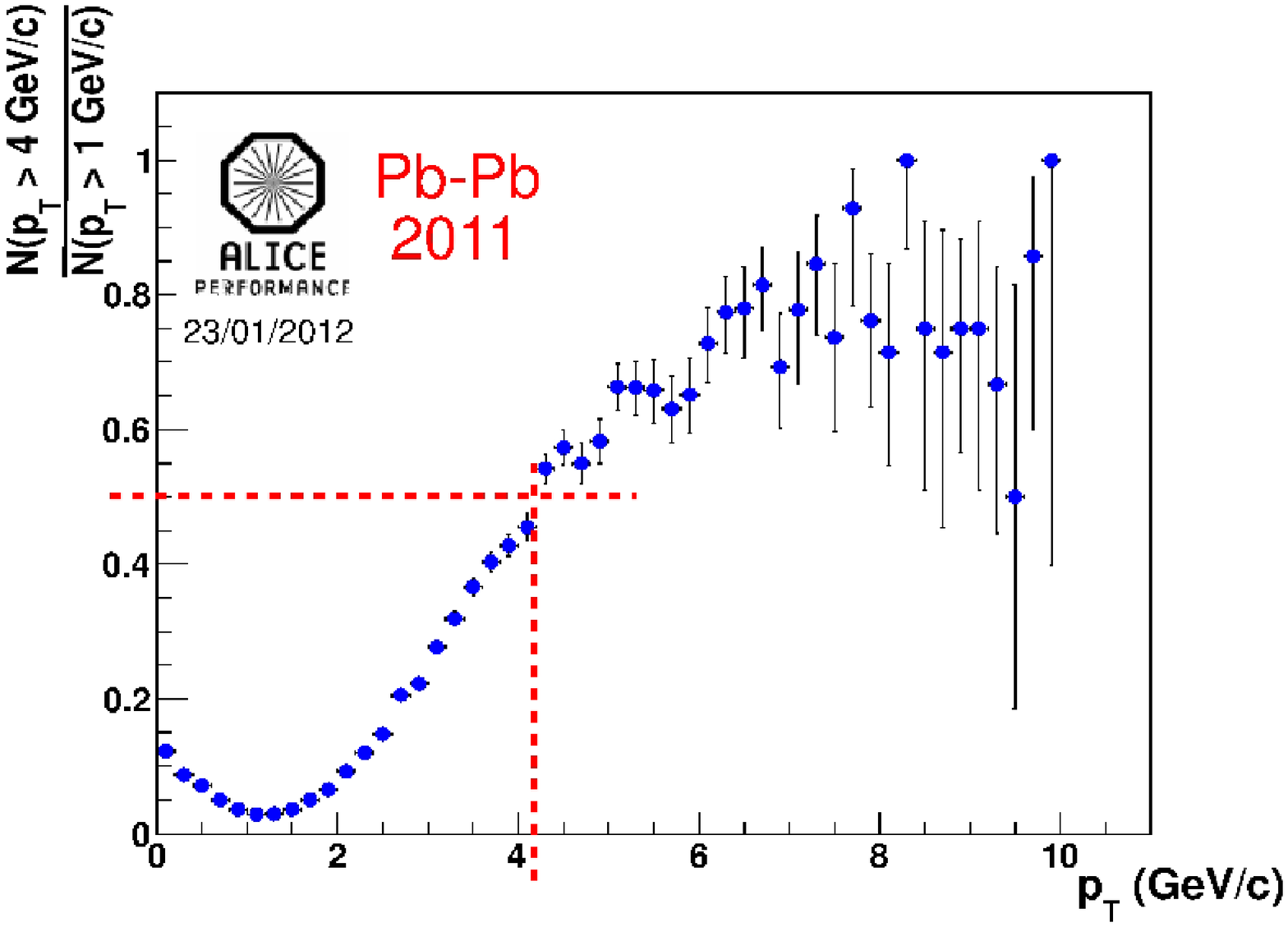}
\caption{Muon trigger turn-on curves in Pb--Pb collisions for the 1~GeV/$c$ threshold  in 2010 (top) and for the 4~GeV/$c$ threshold in 2011 (bottom).}
\label{h_l}
\end{figure}

The $p_{\rm{T}}$ thresholds are fixed by physics and bandwidth considerations. They are defined as the muon $p_{\rm{T}}$ for which an efficiency of 50\% is reached. In 2010 (2011) the low-$p_{\rm{T}}$ threshold was set to 0.5~GeV/$c$ (1~GeV/$c$) and the high-$p_{\rm{T}}$ threshold to 1~GeV/$c$ (4~GeV/$c$). The hardware settings corresponding to the chosen thresholds are determined by simulation with muons of known transverse momentum~\cite{simulaz}.  In real data, the nominal high-$p_{\rm{T}}$ thresholds can be verified by measuring the trigger turn-on curves. The curves are obtained by the ratio of the reconstructed muon $p_{\rm{T}}$ distribution in the high-$p_{\rm{T}}$ muon sample to the reconstructed muon $p_{\rm{T}}$ distribution in the low-$p_{\rm{T}}$ sample. The reconstructed $p_{\rm{T}}$ is measured  by the muon tracking system.
The measured ratios for Pb--Pb collisions in 2010 and 2011 are shown in figure \ref{h_l}.

A plateau with saturation value close to unity is seen for large $p_{\rm{T}}$ values. As expected, the $p_{\rm{T}}$ values corresponding to a value of the ratio of 0.5 are close to the requested values of 1~GeV/$c$ and 4~GeV/$c$. The behaviour of the ratios at very low $p_{\rm{T}}$ can be attributed to the fact that low-$p_{\rm{T}}$ muons passing through the muon filter\footnote{The muon filter is a 1.2~m (7 hadronic interaction lenghts) thick iron absorber  located between the tracking and the trigger system.} placed between the two subdetectors can be affected by multiple scattering effects causing them to be flagged as high-$p_{\rm{T}}$ tracks by the trigger algorithm.

\subsection{Trigger selectivity}

\begin{figure}[htbp]
\centering
\includegraphics[width=0.55\textwidth]{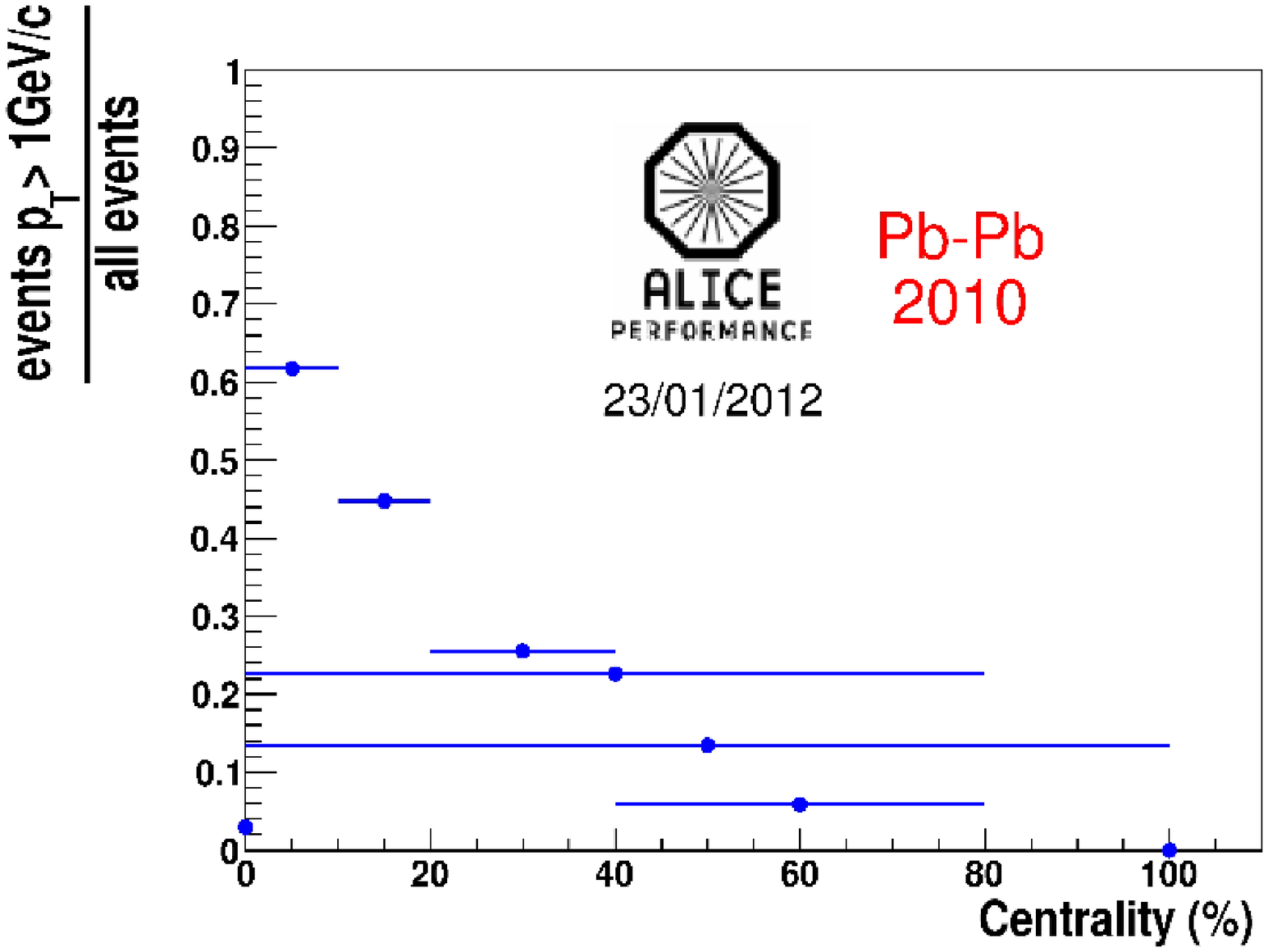}
\includegraphics[width=0.55\textwidth]{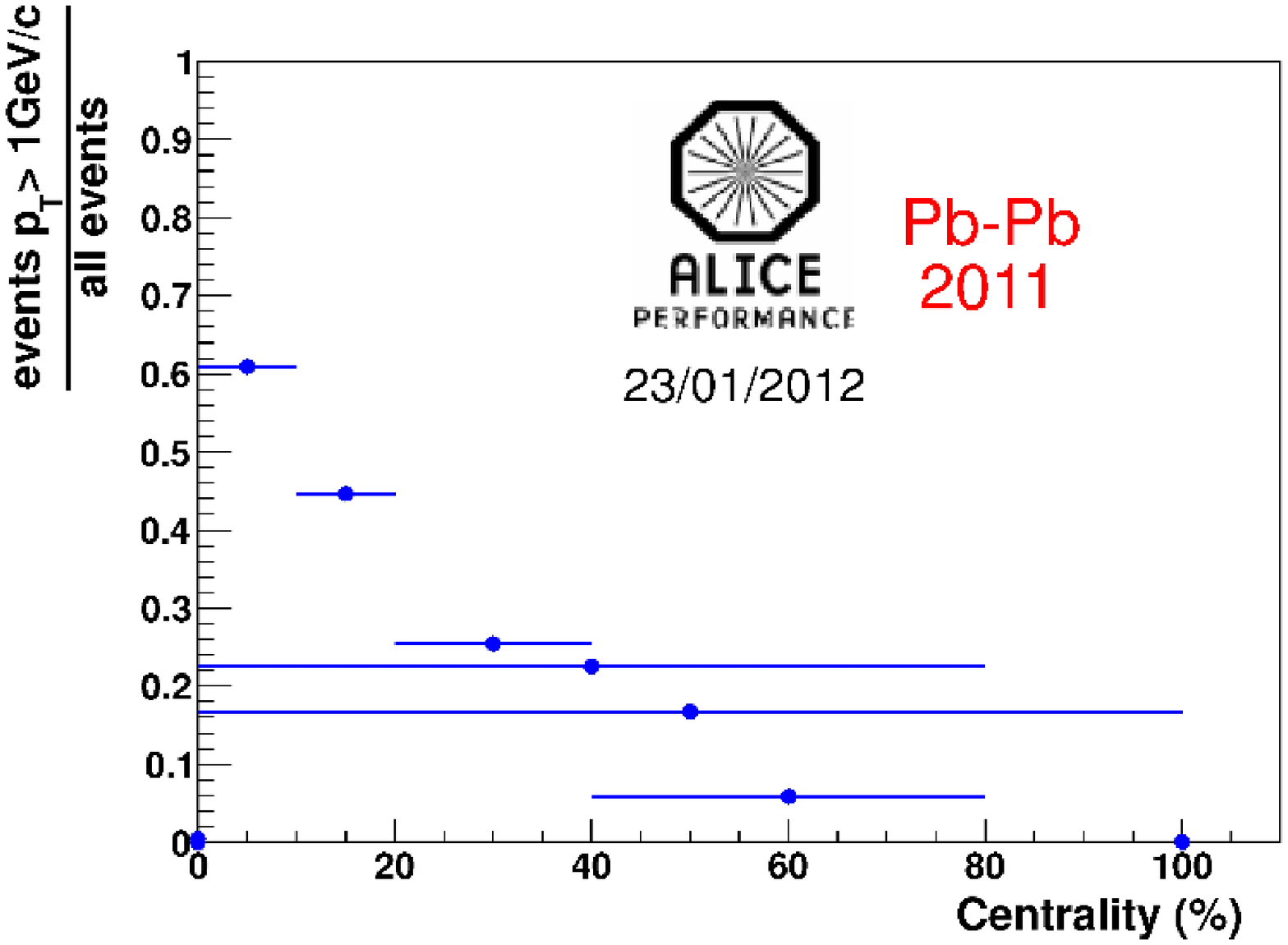}
\includegraphics[width=0.55\textwidth]{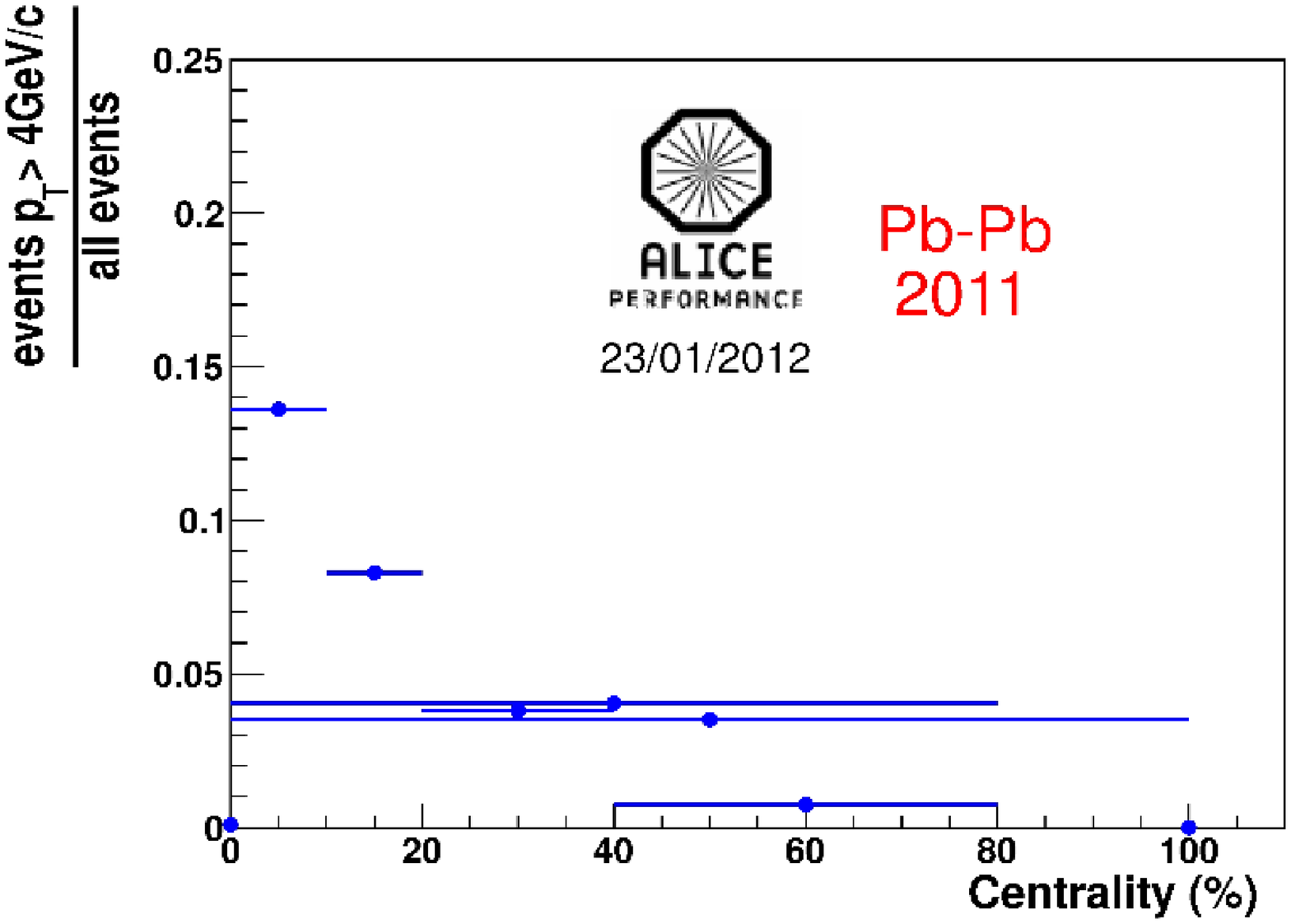}
\caption{Muon trigger selectivities in Pb--Pb collisions in 2010 and 2011, with $p_{\rm{T}}$ thresholds of 1~GeV/$c$ (top, center) and 4~GeV/$c$ (bottom). The horizontal error bars represent the width of the centrality bin.}
\label{global}
\end{figure}

Figure \ref{global} shows the trigger selectivity as a function of centrality, for different $p_{\rm{T}}$ thresholds.  The trigger selectivity is defined as the ratio between the number of minimum bias events containing at least one muon with $p_{\rm{T}}$ larger than the threshold  and the total number of events within a given centrality range.

As expected, the selectivity is lower in the most central classes, where, as shown in~\ref{subsec:mult}, muon production is enhanced by the large number of nucleon-nucleon collisions.
In table \ref{selectivity} the trigger selectivities for two centrality bins are reported. The agreement between the 2010 and 2011 ratios with the same threshold is satisfactory; the small (1.5\%) difference seen between the 2010 and 2011 selectivities in the most central collisions can be ascribed to slight variations in the fraction of  unavailable channels or the RPC efficiency, affecting the overall trigger efficiency. 

\begin{table}[htbp]
\begin{center}
\caption{Trigger selectivity for two centrality ranges and different thresholds.}
\label{selectivity}
\begin{tabular}{|c|c|c|c|}
\cline{2-4}
\multicolumn{1}{c|}{} & \textbf{2010}    & \textbf{2011}    & \textbf{2011} \\
\multicolumn{1}{c|}{} & $p_{\rm{T}}$~>~1~GeV/$c$ & $p_{\rm{T}}$~>~1~GeV/$c$ & $p_{\rm{T}}$~>~4~GeV/$c$ \\
\hline
\textbf{0\%-10\%}  & (61.8$\pm$0.1)\% & (60.9$\pm$0.1)\% & (13.61$\pm$0.05)\% \\
\hline
\textbf{40\%-80\%} & ~~(5.81$\pm$0.02)\% & ~~(5.88$\pm$0.02)\% & ~~(0.73$\pm$0.01)\% \\
\hline
\end{tabular}
\end{center}
\end{table}

\subsection{Muon tracking-trigger matching}

\begin{figure}[htbp]
\centering
\includegraphics[width=0.55\textwidth]{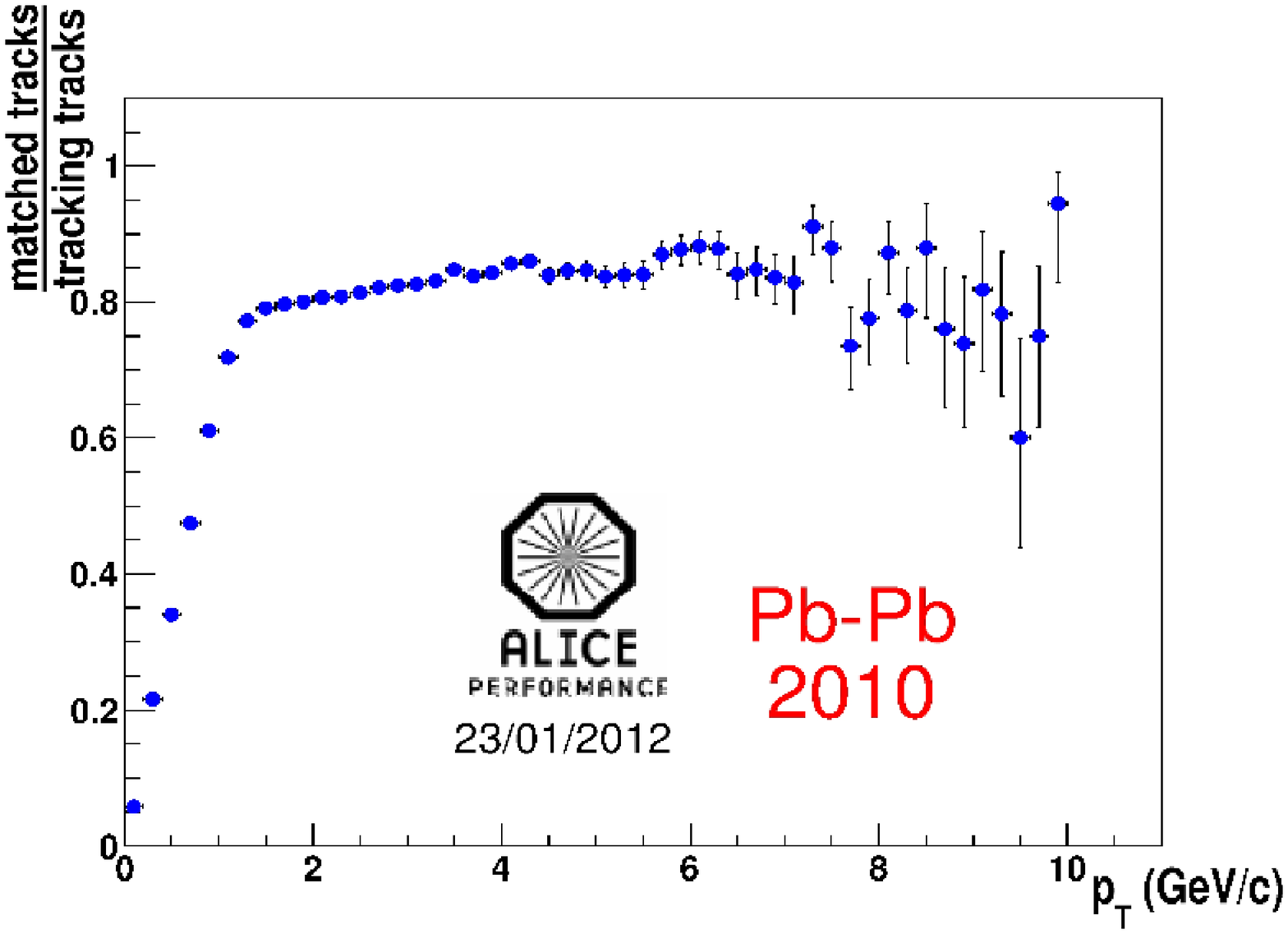}
\includegraphics[width=0.55\textwidth]{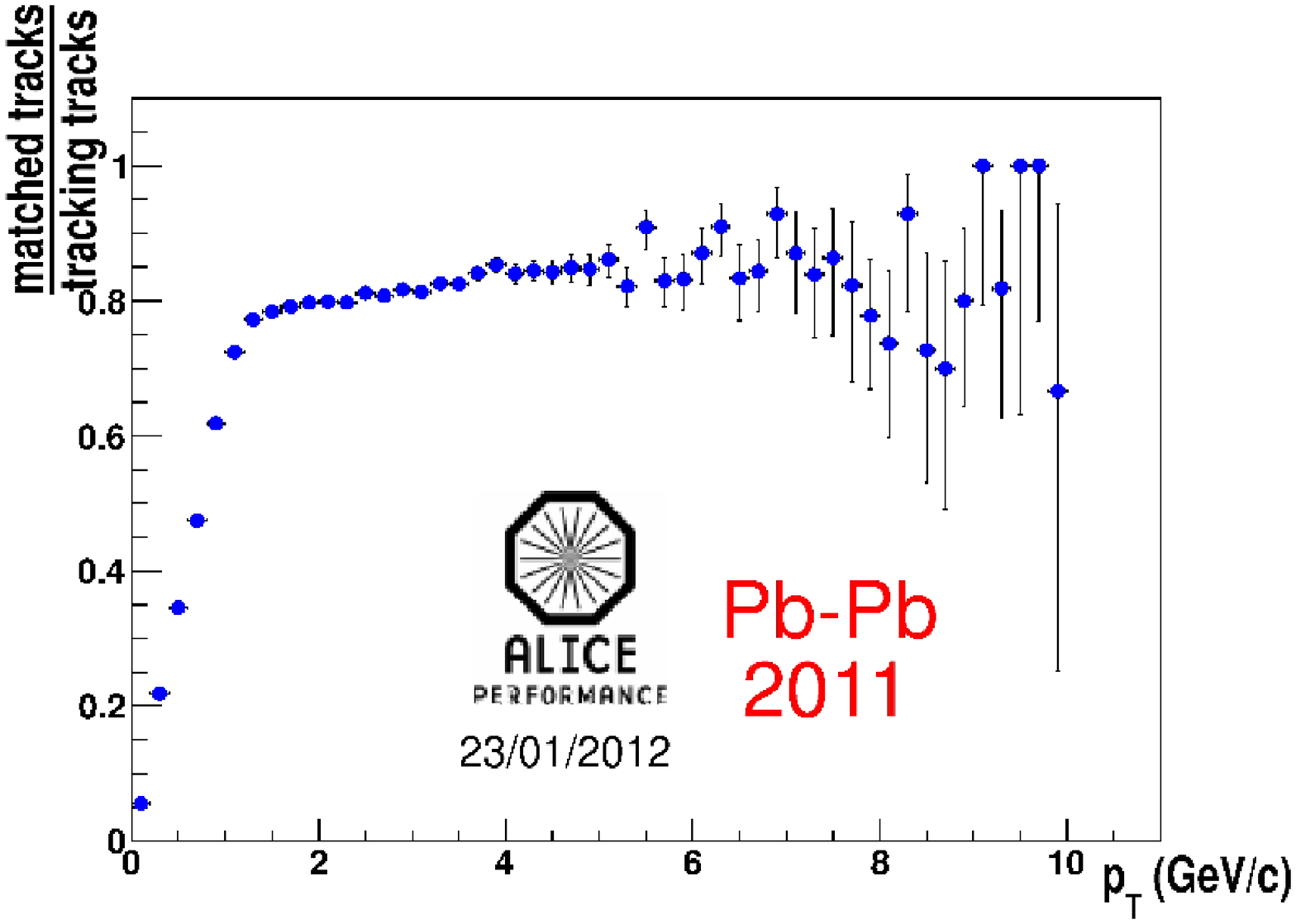}
\caption{Number of matched tracks over the number of tracks detected at least by
         the muon tracking as a function of $p_{\rm{T}}$ for 2010 (top) and 2011 (bottom)
         Pb--Pb collisions.}
\label{match}
\end{figure}

Figure~\ref{match} shows the ratio of the number of tracks detected by both the muon tracking and the muon trigger system (matched tracks) to the number of tracks detected by the tracking system, as a function of $p_{\rm{T}}$, for minimum bias events, in 2010 (top) and 2011 (bottom).

The results obtained for the two years are very similar.
We note that the ratio does not reach unity, the saturation value at high $p_{\rm{T}}$ being $\simeq$0.85. According to Pb--Pb collision  simulations performed with the HIJING event generator, the fraction of unmatched tracks is largely accounted for by the contamination of  hadrons in the sample of reconstructed tracks: these are stopped in the muon filter and are not detected by the trigger system. Other effects, such as trigger and matching inefficiences and tracks from beam-gas interactions, play a marginal role in defining the value of the ratio.

\section{Conclusions}

The ALICE muon trigger system has been fully operational during the first two years of data-taking at the LHC. The observed RPC performance are in agreement with the design values. The average RPC efficiency is about 95\%; the cluster size is 1.4 with 2~cm wide strips; the dark current is about 0.1~nA/cm$^2$ and the dark counting rate is about 0.05~Hz/cm$^2$. The values of all these parameters are remarkably stable in time, with the sole exception of the dark current. 

From the analysis of 2010 and 2011 muon trigger data in Pb--Pb collisions it is possible to conclude that: the muon trigger system has shown a stable behavior; the trigger decision algorithm is reliable and selective; the muon trigger is very efficient in  rejecting hadrons detected by the muon spectrometer tracking system: it actually acts as a muon identifier.

The ALICE muon spectrometer is playing a crucial role in the ALICE physics program and, also in virtue of the performance and stability of the muon trigger system, is expected to continue to do so in the next years of LHC operation.

%During 2011 Pb--Pb collisions ALICE collected an integrated luminosity of
%$\sim$144 $\mu$b$^{-1}$ (15 times more than in 2010) with $\mathcal{L}_{max}$ = 5 10$^{26}$~Hz/cm$^2$
%(an order of magnitude higher with respect to 2010).
%The maximum collision rate in 2011 was of $\sim$4 kHz corresponding to $\sim$600~Hz
%of single muon trigger above 1~GeV/$c$.
%With these numbers and also thanks to the good muon trigger performance, the muon
%spectrometer reconstructed in Pb--Pb 2010 more than 2500 J/$\psi$ \cite{risultati}
%and in Pb--Pb 2011 about 40000 J/$\psi$ and more than 100 $\Upsilon$ are foreseen.

%\acknowledgments

\end{document}